\makeatletter
\documentstyle[11pt,epsfig]{article}
\newcounter{subequation}[equation]

\makeatletter

\def\thesubequation{\theequation\@alph\c@subequation}
\def\@subeqnnum{{\rm (\thesubequation)}}
\def\slabel#1{\@bsphack\if@filesw {\let\thepage\relax
   \xdef\@gtempa{\write\@auxout{\string
      \newlabel{#1}{{\thesubequation}{\thepage}}}}}\@gtempa
   \if@nobreak \ifvmode\nobreak\fi\fi\fi\@esphack}
\def\subeqnarray{\stepcounter{equation}
\let\@currentlabel=\theequation\global\c@subequation\@ne
\global\@eqnswtrue
\global\@eqcnt\z@\tabskip\@centering\let\\=\@subeqncr
$$\halign to \displaywidth\bgroup\@eqnsel\hskip\@centering
  $\displaystyle\tabskip\z@{##}$&\global\@eqcnt\@ne
  \hskip 2\arraycolsep \hfil${##}$\hfil
  &\global\@eqcnt\tw@ \hskip 2\arraycolsep
  $\displaystyle\tabskip\z@{##}$\hfil
   \tabskip\@centering&\llap{##}\tabskip\z@\cr}
\def\endsubeqnarray{\@@subeqncr\egroup
                     $$\global\@ignoretrue}
\def\@subeqncr{{\ifnum0=`}\fi\@ifstar{\global\@eqpen\@M
    \@ysubeqncr}{\global\@eqpen\interdisplaylinepenalty \@ysubeqncr}}
\def\@ysubeqncr{\@ifnextchar [{\@xsubeqncr}{\@xsubeqncr[\z@]}}
\def\@xsubeqncr[#1]{\ifnum0=`{\fi}\@@subeqncr
   \noalign{\penalty\@eqpen\vskip\jot\vskip #1\relax}}
\def\@@subeqncr{\let\@tempa\relax
    \ifcase\@eqcnt \def\@tempa{& & &}\or \def\@tempa{& &}
      \else \def\@tempa{&}\fi
     \@tempa \if@eqnsw\@subeqnnum\refstepcounter{subequation}\fi
     \global\@eqnswtrue\global\@eqcnt\z@\cr}
\let\@ssubeqncr=\@subeqncr
\@namedef{subeqnarray*}{\def\@subeqncr{\nonumber\@ssubeqncr}\subeqnarray}
\@namedef{endsubeqnarray*}{\global\advance\c@equation\m@ne%
                           \nonumber\endsubeqnarray}

\makeatletter
\@addtoreset{equation}{section}
\makeatother
\renewcommand{\theequation}{\thesection.\arabic{equation}}

\def\dalemb#1#2{{\vbox{\hrule height .#2pt
        \hbox{\vrule width.#2pt height#1pt \kern#1pt
                \vrule width.#2pt}
        \hrule height.#2pt}}}

    \let\e=\epsilon
  \let\q=\theta  
  \let\n=\nu

 \def\bd{\begin{document}} \def\ed{\end{document}}
\def\ds{\documentstyle} \let\fr=\frac \let\bl=\bigl \let\br=\bigr
\let\Br=\Bigr \let\Bl=\Bigl
\let\bm=\bibitem
\let\na=\nabla
\let\pa=\partial \let\ov=\overline
\def\ie{{\it i.e.\ }}
\newcommand{\be}{\begin{equation}}
\newcommand{\ee}{\end{equation}}
\def\ba{\begin{array}}
\def\ea{\end{array}}
\def\ft#1#2{{\textstyle{{\scriptstyle #1}\over {\scriptstyle #2}}}}
\def\fft#1#2{{#1 \over #2}}
\def\del{\partial}
\def\sst#1{{\scriptscriptstyle #1}}
\def\oneone{\rlap 1\mkern4mu{\rm l}}
\def\e7{E_{7(+7)}}
\def\td{\tilde}
\def\wtd{\widetilde}
\def\im{{\rm i}}
\def\bog{Bogomol'nyi\ }
\def\q{{\tilde q}}
\def\hast{{\hat\ast}}
\def\0{{\sst{(0)}}}
\def\1{{\sst{(1)}}}
\def\2{{\sst{(2)}}}
\def\3{{\sst{(3)}}}
\def\4{{\sst{(4)}}}
\def\5{{\sst{(5)}}}
\def\6{{\sst{(6)}}}
\def\7{{\sst{(7)}}}
\def\8{{\sst{(8)}}}
\def\n{{\sst{(n)}}}
\def\oo{{\"o}}
\def\hA{\hat{\cal A}}
\def\ns{{\sst {\rm NS}}}
\def\rr{{\sst {\rm RR}}}
\def\tH{{\widetilde H}}
\def\tB{{\widetilde B}}
\def\cA{{\cal A}}
\def\cF{{\cal F}}
\def\tF{{\wtd F}}
\def\Z{\rlap{\sf Z}\mkern3mu{\sf Z}}
\def\ep{{\epsilon}}
\def\IIA{{\rm IIA}}
\def\IIB{{\rm IIB}}
\def\ads{{\rm AdS}}
\def\R{\rlap{\rm I}\mkern3mu{\rm R}}
\def\mapright#1{\smash{\mathop{-\!\!\!-\!\!\!-\!\!\!-\!\!\!-\!\!\!
             \longrightarrow}\limits^{#1}}}

\def\Ei{{\hbox{Ei}}}
\def\Ci{{\hbox{Ci}}}
\def\Si{{\hbox{Si}}}

\newcommand{\ho}[1]{$\, ^{#1}$}
\newcommand{\hoch}[1]{$\, ^{#1}$}
\newcommand{\bea}{\begin{eqnarray}}
\newcommand{\eea}{\end{eqnarray}}
\newcommand{\ra}{\rightarrow}
\newcommand{\lra}{\longrightarrow}
\newcommand{\Lra}{\Leftrightarrow}
\newcommand{\aap}{\alpha^\prime}
\newcommand{\bp}{\tilde \beta^\prime}
\newcommand{\tr}{{\rm tr} }
\newcommand{\Tr}{{\rm Tr} }
\newcommand{\NP}{Nucl. Phys. }
\newcommand{\tamphys}{\it Center for Theoretical Physics,
Texas A\&M University, College Station, TX 77843}
\newcommand{\upenn}{\it Dept. of Phys. and Astro.,
University of Pennsylvania,
Philadelphia, PA 19104}

\newcommand{\auth}{J. F. V\'azquez-Poritz}

\begin{document}
\begin{flushright}
UPR/920-T \\
January 2001\\
\hfill{\bf hep-th/0101057}\\
\end{flushright}


\begin{center}

{\large {\bf Localized Massive Gravity from B fields}
}

\vspace{20pt}

\auth

\vspace{10pt}
{\upenn}

\vspace{30pt}

\underline{ABSTRACT}
\end{center}

We consider five-dimensional domain-wall solutions which arise from a
sphere reduction in M-theory or string theory and have the 
higher-dimensional interpretation of the near-horizon region of various
p-branes in constant, background B fields. We analyze the fluctuation
spectrum of linearized gravity and find that there is a massive state
which is localized and plays the role of the four-dimensional graviton.

{\vfill\leftline{}\vfill
\vskip 10pt \footnoterule {\footnotesize Research supported
in part by DOE grant DOE-FG02-95ER40893
\vskip  -12pt} 

\pagebreak
\setcounter{page}{1}


\section{Introduction}

Randall and Sundrum \cite{randall1,randall2} have shown that, with fine
tuned brane tension, a flat 3-brane embedded in $AdS_5$ can have a single,
massless bound state. Four-dimensional gravity is recovered at low-energy
scales. It has also been proposed that part or all of gravitational
interactions are the result of massive gravitons. For example, in one
model, gravitational interactions are due to the net effect of the
massless graviton and ultra-light Kaluza-Klein 
state(s) \cite{kogan1,mous,kogan2,kogan3}. In another proposal, there is
no normalizable massless graviton and four-dimensional gravity is
reproduced at intermediate scales from a resonance-like behavior of the
Kaluza-Klein wave functions \cite{kogan2,kogan3,gregory,csaki2,dvali}.

Recently, it has been shown that an $AdS_4$ brane in $AdS_5$ does not have
a normalizable massless graviton. Instead, there is a very light, but
massive bound state graviton mode, which reproduces four-dimensional
gravity \cite{karch,kogan,kogan4,porrati1}. The bound state mass as a
function of brane
tension, as well as the modified law of gravity, were explored in
\cite{schwartz,miemiec}.

A brane world in which two extra infinite spatial dimensions do not
commute has recently been used in order to localize scalar fields as well
as fermionic and gauge fields \cite{pilo}. Also, the cosmological
evolution of the four-dimensional universe on the probe D$3$-brane in
geodesic motion in the curved background of the source D$p$-brane with
nonzero NS $B$ field has been explored \cite{youm}. 

In this paper, we consider brane world models which arise from a sphere
reduction in M-theory or string theory, as the near-horizon of $p$-branes
\cite{cvetic} with a constant, background $B$-field on the
world-volume. The dual field theory is non-commutative Yang-Mills
\cite{itzhaki,maldacena}. 

We consider string theoretic $p$-brane solutions for $p=3,4,5$
with $0,1,2$ world-volume dimensions wrapped around a compact manifold,
respectively. For $p>5$, the space of the extra large dimension has
finite-volume, with a naked singularity at the end, for which case the
localization of gravity is trivial. 

For all of the cases we study, we find that there is no normalizable
massless graviton, but there is a massive bound state graviton which plays
the role of the four-dimensional graviton. This yields another class of
examples for which gravitational interactions are entirely the result of
massive gravitons. As we will see, the bound mass increases continuously
from zero as a constant, background $B$ or $E$-field is turned on, except 
for the case $p=5$, for which there is a mass gap. 

This paper is organized as follows. In section 2, we find the graviton
wave equation in the background of five-dimensional domain-walls that
originate in string theory as $p$-branes in constant background $B$ and
$E$-fields. In section 3, we show that the massless graviton wave function
is not normalizable for nonzero background $B$ or $E$ fields, indicating
that the massless graviton is not localized on the brane. In section 4, we
show that there is a massive graviton localized on the brane for nonzero
$B$ and $E$-fields. Lastly, an appendix shows how the graviton wave
equation can be solved exactly for $p=5$.

\section{The graviton wave equation}
\subsection{With background B field}

Consider the metric of a $p$-brane (expressed in the string frame)
in a constant NS $B$ field in the $2$,$3$ directions \cite{maldacena,oz}:
\begin{eqnarray}
ds_{10}^2 &=& H^{-1/2}\big(
-dt^2+dx_1^2+h(dx_2^2+dx_3^2)+dy_i^2\big) +l_s^4 H^{1/2}(du^2+u^2
d\Omega_5^2),\\ \nonumber H &=& 1+\frac{R^{7-p}}{l_s^4 u^{7-p}},\
\ \ \ \ \ h^{-1}=H^{-1} \sin^2\theta+\cos^2\theta,\label{metric}
\end{eqnarray}
where $u=r/l_s^2$ is a dimensionless radial parameter and
$i=1,..,p-3$. The $B$ field and dilaton are given by
\be 
B_{23}=\tan \theta H^{-1}h,\ \ \ \ \ e^{2\varphi}=g^2
H^{(3-p)/2}h,
\ee
\be 
\cos \theta R^{7-p}=(4\pi)^{(5-p)/2}\Gamma{\Big( \frac{7-p}{2}
\Big) } g_s l_s^{p-3} N, 
\ee
where $N$ is the number of $p$-branes and $g_s$ is the asymptotic
value of the coupling constant. $\theta$ is known as the non-commutativity 
parameter, and is related to the asymptotic value of the
$B$-field: $B_{23}^{\infty}=\tan \theta$.

Consider the following decoupling limit, in which we take the $B$ field to
infinity \cite{maldacena}:
\begin{eqnarray}
l_s \rightarrow 0,\ \ \ \ \ \ \ l_s^2 \tan \theta =b,
\\ \nonumber \tilde{x}_{2,3}=\frac{b}{l_s^2} x_{2,3},
\end{eqnarray}
where $b,u,\tilde{x}_{\mu}$ and $g_s l_s^{p-5}$ stay
fixed. Thus,
\be
l_s^{-2} ds^2=\big( \frac{u}{R}\big)^{(7-p)/2}
\big(-dt^2+dx_1^2+\tilde{h}(d\tilde{x}_2^2+d\tilde{x}_3^2)+dy_i^2\big)+\big(
\frac{R}{u}\big)^{(7-p)/2} (du^2+u^2 d\Omega_{8-p}^2),
\ee
where $\tilde{h}^{-1}=1+b^2 (u/R)^{7-p}$.
The dual field theory is Yang-Mills with noncommuting $2,3$
coordinates \cite{maldacena}. For small $u$, the above solution reduces to
$AdS_5 \times S^5$, corresponding to ordinary Yang-Mills living in the
IR region of the dual field theory, which is also the case if we had
taken the decoupling limit with finite $B$-field. From this point
on, we drop the $\tilde{}$ on the coordinates. For $p \ne 5$, we change
coordinates to $u/R \equiv (1+k|z|)^{2/(p-5)}$ \cite{cvetic}:
\be
l_s^{-2} ds^2=(1+k|z|)^{\frac{7-p}{p-5}}\big(-dt^2+dx_1^2+ 
\tilde{h}(dx_2^2+dx_3^2)+dy_i^2 +dz^2\big)+(1+k|z|)^{\frac{p-3}{p-5}}
d\Omega_{8-p}^2,
\ee
For $p=5$, we change coordinates to $u/R \equiv e^{-k|z|/2}$
\cite{cvetic}:
\be
l_s^{-2} ds^2=e^{-k|z|/2}\big(-dt^2+dx_1^2+\tilde{h}(dx_2^2+dx_3^2)+
dy_i^2 +dz^2+d\Omega_3^2\big).
\ee
After dimensional-reduction over $S^{8-p}$, the above metric corresponds
to a $p+2$-dimensional domain wall at $z=0$. We will consider the case
where $y_i$ are wrapped around a compact manifold, so that there is a
$1+3$-dimensional dual field theory on the remaining world-volume
coordinates. The equation of motion for the graviton fluctuation
$\Phi=g^{00}h_{01}$ is
\be
\partial_M \sqrt{-g}e^{-2\varphi}g^{MN}\partial_N \Phi=0,
\label{eom}
\ee
where $h_{01}$ is associated with the energy-momentum tensor
component $T_{01}$ of the Yang-Mills theory. For $p \ne 5$, consider
the ansatz
\be
\Phi=\phi(z)e^{ip \cdot x}=(1+k|z|)^{\frac{9-p}{2(5-p)}}\psi(z)e^{ip
\cdot x}. \label{ansatz}
\ee
Thus,
\be
-\psi^{''}+U\psi=m^2 \psi, \label{wave}
\ee
\be
U=\frac{(p-9)(3p-19)k^2}{4(5-p)^2(1+k|z|)^2}+\frac{p-9}{5-p}k\delta(z)+
\frac{\alpha^2}{(1+k|z|)^{2(7-p)}{5-p}},
\label{U}
\ee
where
\be
\alpha \equiv b \sqrt{(p_2^2+p_3^2)},
\ee
and $m^2=-p^2$. 

For $p=5$, consider the ansatz
\be
\Phi=\phi(z)e^{ip \cdot x}=e^{k|z|/2}\psi(z)e^{ip\cdot x}. \label{a5}
\ee
Thus, $\psi$ satisfies (\ref{wave}) with
\be
U=\frac{1}{4}k^2-k\delta(z)+\alpha^2 e^{-k|z|}.
\label{u5}
\ee
For more general directions of the world-volume $B$ field, $\alpha
\equiv b \sqrt{p_i^2}$, where $p_i$ are the momenta along the large,
non-commuting directions in the world-volume. Thus, if only the wrapped
$y_i$ coordinates are  non-commuting, then $\alpha=0$. 

\subsection{With background B and E fields}

Consider the metric of a $p$-brane in a constant two-component
NS $B$ field background \cite{maldacena,oz}:
\begin{eqnarray}
ds^2 &=& H^{-1/2}\big(
h_e(-dt^2+dx_1^2)+h_m(dx_2^2+dx_3^2)+dy_i^2\big)
+l_s^4 H^{1/2}(du^2+u^2 d\Omega_{8-p}^2),\\ \nonumber
H &=& 1+\frac{R^{7-p}}{l_s^4 u^{7-p}},\ \ \ \ h_m^{-1}=H^{-1}
\sin^2\theta_m+\cos^2\theta_m,\ \ \ \ h_e^{-1}=-H^{-1}\sinh^2
\theta_e+\cosh^2 \theta_e.
\end{eqnarray}
The $B$ and $E$ field components and dilaton are given by
\be
B_{23}=\tan \theta_m H^{-1}h_m,\ \ \ \ \ E_{01}=-\tanh \theta_e
H^{-1} h_e,\ \ \ \ \ e^{2\varphi}=g^2 H^{\frac{3-p}{2}}h_e h_m.
\ee
Consider the following decoupling limit \cite{maldacena}:
\begin{eqnarray}
l_s \rightarrow 0,\ \ \ \ \ \cosh \theta_e=
\frac{b^{'}}{l_s},\ \ \ \ \ \cos \theta_m=\rm{fixed},\\
\nonumber x_{0,1}=\frac{b^{'}}{l_s^2}\tilde{x}_{0,1},
\ \ \ x_{2,3}=l_s \cos \theta_m \tilde{x}_{2,3},
\end{eqnarray}
where $b^{'},u,\tilde{x}_{\mu}$ and $g_s l_s^{p-7}$ remain fixed. Thus,
\be
l_s^{-2} ds^2=H^{1/2}\Big( \big(\frac{u}{R}\big)^{7-p}
(-d\tilde{t}^2+d\tilde{x}_1^2)+\big(\frac{u}{R}\big)^{7-p}
\tilde{h}_m(d\tilde{x}_2^2+d\tilde{x}_3^2)+H^{-1}dy_i^2+du^2+
u^2 d\Omega_{8-p}^2\Big),
\ee
where we define
\be
\tilde{h}_m^{-1}=1+\big(\frac{u}{R}\big)^{7-p}\cos^{-2} \theta_m.
\ee
The dual field theory is Yang-Mills with noncommuting $0,1$ and
$2,3$ coordinates \cite{maldacena}. Once again, we drop the $\tilde{}$ on
the coordinates. For $p \ne 5$, we change coordinates to $u/R \equiv
(1+k|z|)^{2/(p-5)}$:
\be
l_s^{-2}ds^2=H^{1/2}\Big((1+k|z|)^{\frac{2(p-7)}{5-p}}\big(-dt^2+dx_1^2+
\tilde{h}_m(dx_2^2+dx_3^2)+dz^2\big)+H^{-1}dy_i^2+
(1+k|z|)^{\frac{4}{p-5}}d\Omega_{8-p}^2 \Big).
\ee
For $p=5$, we change coordinates to $u/R \equiv e^{-k|z|/2}$:
\be
l_s^{-2}ds^2=H^{1/2} \big( e^{-k|z|} (-dt^2+dx_1^2+
\tilde{h}_m(dx_2^2+dx_3^2)+dz^2+d\Omega_3^2)+H^{-1}dy_i^2\big).
\ee
As in the case of a one-component $B$-field, we consider a
dimensional-reduction over $S^{8-p}$ and $y_i$ wrapped around a
compact manifold. One effect of the $E$-field, as opposed to the
$B$-field, is to add a breathing-mode to $S^{8-p}$. For $p \ne 5$, we
insert this metric into the graviton equation of motion (\ref{eom}) with
the ansatz (\ref{ansatz}) and obtain (\ref{wave}) with $U$ given by
(\ref{U}) and with
\be
\alpha \equiv \frac{\sqrt{p_2^2+p_3^2}}{\cos \theta_m}.
\ee
For $p=5$, we use the wave function ansatz (\ref{a5}) and obtain
(\ref{wave}) with $U$ given by (\ref{u5}). Again, in general $\alpha
\equiv \sqrt{p_i^2}/\cos \theta_m$, where $p_i$ are the momenta along the
large, non-commuting directions. 

In this paper, we will focus on the $p=3$ case, for which the volcano
potential $U$ is plotted in Figure 1, for $\alpha=0$ (solid line) and
$\alpha=2$ (dotted line).
\begin{figure}
   \epsfxsize=4.0in
   \centerline{\epsffile{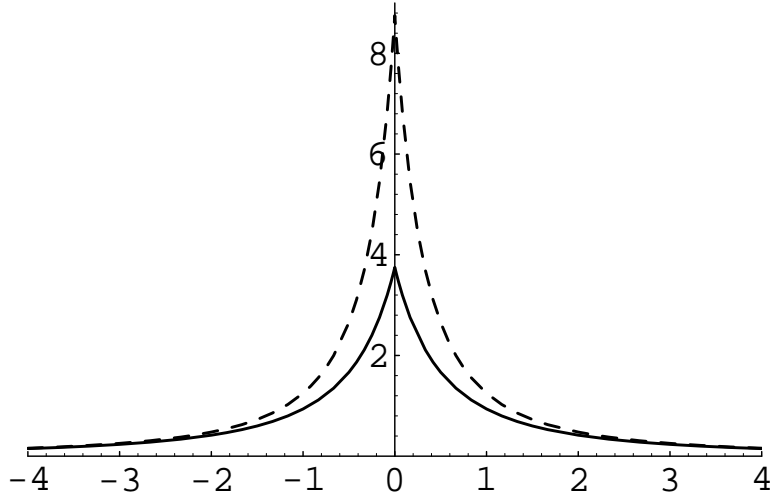}}
   \caption[FIG. \arabic{figure}.]{$U(z)$ for $\alpha=0$ (solid line) and
$\alpha=2$ (dotted line) for $p=3$.}
\end{figure}
As can be seen, the effect of the B and E fields is the raise
the peak of the potential. This is the case in general for $p \le 5$.

\section{The massless graviton}

Consider the coordinate transformation 
\be
u/R \equiv (1+k|w|)^{-1}\label{ztrans}
\ee
and the wave function transformation 
\be
\Theta(w)=(1+k|w|)^{\frac{p-3}{4}}\psi(z).\label{wavetrans}
\label{aa}
\ee
applied to the graviton wave equation (\ref{wave}).
Thus,
\be
-\Theta^{''}+V\Theta=\frac{m^2}{(1+k|w|)^{p-3}}\Theta,\label{Wave} 
\ee
where
\be
V=\frac{(6-p)(8-p)k^2}{4(1+k|w|)^2}-(6-p)k\delta(w)+
\frac{\alpha^2}{(1+k|w|)^4}.
\label{V}
\ee
For the massless case, this is a Schr\"{o}dinger-type equation. The
wave function solution can be written in terms of Bessel functions as
\be
\Theta(w)=N_0 (1+k|w|)^{1/2} \big[ J_{\nu} \Big(\frac{i\alpha}{1+k|w|}
\Big) +A(\alpha) Y_{\nu} \Big( \frac{i \alpha}{1+k|w|}\Big) \big],
\label{solution}
\ee
where $\nu=(7-p)/2$ and $N_0$ is the normalization constant, in the
event that the above solution is normalizable. $A(\alpha)$ is solved by
considering the $\delta (w)$ boundary condition:
\be
A(\alpha)=\frac{2\nu J_{\nu}(i \alpha)+i\alpha[J_{\nu+1}(i\alpha)-
J_{\nu-1}(i\alpha)}{-2\nu Y_{\nu}(i\alpha)+i\alpha[Y_{\nu-1}(i\alpha)-
Y_{\nu+1}(i\alpha)]},\label{A}
\ee
where we used $2J_n^{\prime}(x)=J_{n-1}(x)-J_{n+1}(x),$ and likewise for 
$Y_n^{\prime}(x)$. In order for the massless graviton to be localized on
the brane, the corresponding wave function must be normalizable, i.e.,
$\int_0^{\infty} |\psi(w,m=0)|^2 dw < \infty$. (\ref{solution}) can be
written in the form
\be
\Theta(w)=(normalizable\ part)+A(non-renormalizable\ part),
\ee
as can be seen from the asymptotic forms of the Bessel functions. Thus,
the massless graviton is only localized on the brane if $A=0$. In this
limit, there are no background B or E fields and we recover the
Randall-Sundrum model.

As shown in Figure 2, for the case of the D$3$-brane, as $\alpha$
increases, the modulus of $A$ approaches unity.
\begin{figure}
   \epsfxsize=4.0in
   \centerline{\epsffile{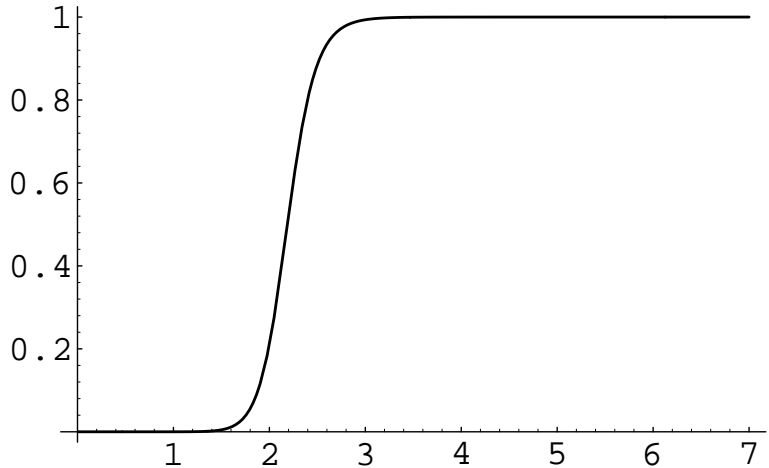}}
   \caption[FIG. \arabic{figure}.]{$|A|^2$ versus $\alpha$ for
$p=3$.}
\end{figure}
$A$ has the same characteristics for $p \le 5$ in general. The issue
arises as to whether there is a massive graviton state bound to the
brane for nonzero $A$. 

\section{Localization of the massive graviton}

\begin{figure}
   \epsfxsize=4.0in
   \centerline{\epsffile{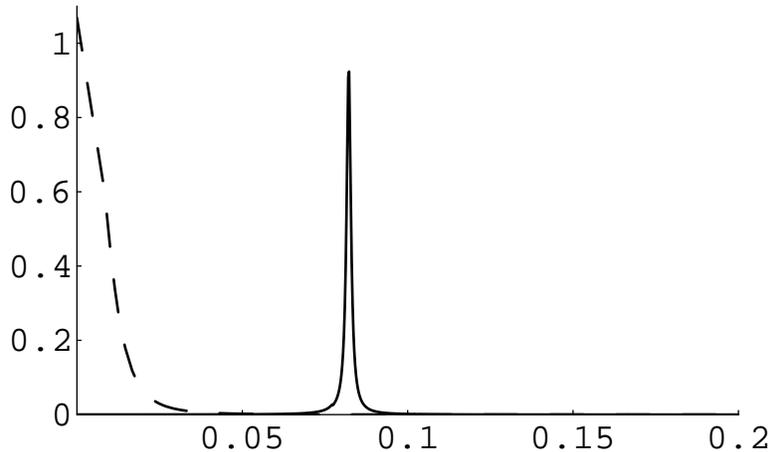}}
   \caption[FIG. \arabic{figure}.]{$|\psi (z=0)|^2$ versus $m$ for
$\alpha=0$ (dotted line) and $\alpha=.02$ (solid line) for $p=3$.}
\end{figure}
We will first concentrate on the case of the brane world as the
dimensionally-reduced near-horizon of the D$3$-brane. For this case, the
massive wave equation  (\ref{wave}) with $U$ given by (\ref{U}) is a
modified Mathieu's equation \cite{hashimoto}, supplemented by the $\delta
(w)$ boundary condition. The exact solution of the modified Mathieu's
equation is known, and has been applied to absorption by $D3$-branes and 
six-dimensional dyonic strings 
\cite{gubser,cvetic1,poritz1,poritz2,park}. However, it is
rather laborious to use the exact solution in this context and so we
content ourselves with plotting numerical solutions-- except for the case
$p=5$, for which the exact solution is a sum of Bessel functions (see
appendix A). 

In order to solve for the unnormalized wave function, we input the
$\delta(z)$ boundary condition and solve the Schr\"{o}dinger-type equation
outwards. We then numerically integrate in order to find the correct
normalization factor.

\begin{figure}
   \epsfxsize=4.0in
   \centerline{\epsffile{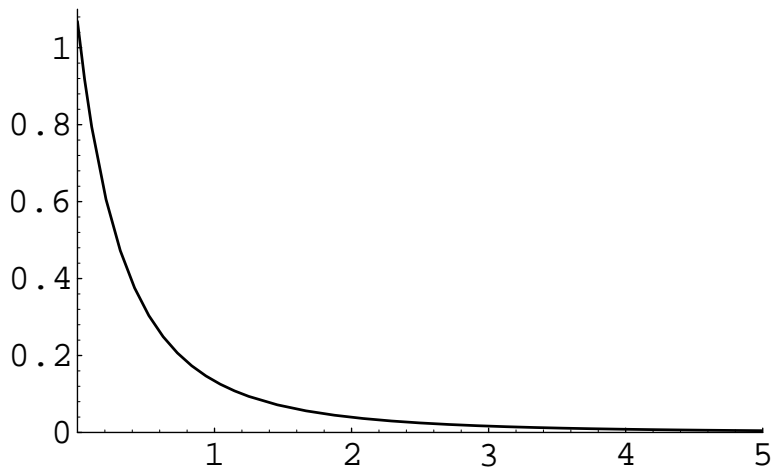}}
   \caption[FIG. \arabic{figure}.]{$|\psi(z)|^2$ for $m/k=0$ for
$p=3$.}
\end{figure}
\begin{figure}
   \epsfxsize=4.0in
   \centerline{\epsffile{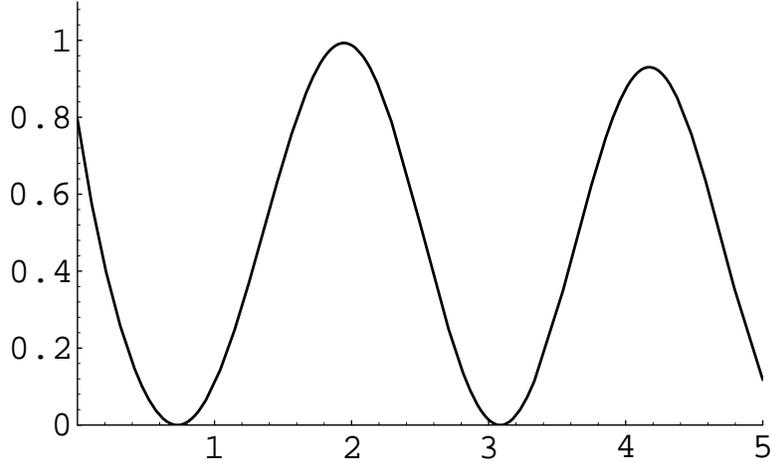}}
   \caption[FIG. \arabic{figure}.]{$|\psi(z)|^2$ for $m/k=1.5$ for
$p=3$.}
\end{figure}
As can be seen from the dotted line in Figure 3, for the case of the
D$3$-brane with no background B and E fields, there is a resonance in the
modulus of the wave function on the brane for $m=0$, which implies that the
massless graviton is localized on the brane. This behavior can also be seen
from Figure 2, by the fact that $A=0$ and the massless solution is
normalizable. 

We may arrive at the same conclusion via Figures 4 and 5, where the
massless wavefunction has a peak on the brane at $z=0$ whereas the 
wavefunction of mass $m=1.5 k$ oscillates without feeling the presence of
the brane \cite{karch}.

In the case of nonzero B or E fields, the resonance in the modulus of the
wave function on the brane is at a nonzero mass. Thus, for a nonzero,
constant background B or E field, a massive graviton is localized on the
brane. This behavior is shown by the solid line in Figure 3, where the
$|\psi(z=0)|^2$ is plotted for $\alpha=.02$, and there is a large
resonance in the wave function on the brane for $m/k=.0822$. Again, this
behavior may also be ascertained from plots of $|\psi(z)|^2$ for various
values of $m/k$. For a mass significantly less than the resonance
mass, the wave function has a peak at $z=0$ that is not relatively high,
while a wave function with a mass significantly greater than the resonance
mass oscillates without feeling the brane.

It is straight-forward to repeat the above analysis for $p=4,5$. Note that
a D$4$-brane reduced on $S^1 \times S^4$ yields the same graviton wave
equation as a M$5$-brane reduced on $T^2 \times S^4$. Likewise, the
D$5$-brane reduced on $T^4 \times S^3$ and a M$5$-brane intersection
reduced on $T^4 \times S^2$ or K3 $\times S^2$ yield the same graviton
wave equation. This was initially observed in \cite{cvetic} in the case of
zero $B$-field.
\begin{figure}
   \epsfxsize=4.0in
   \centerline{\epsffile{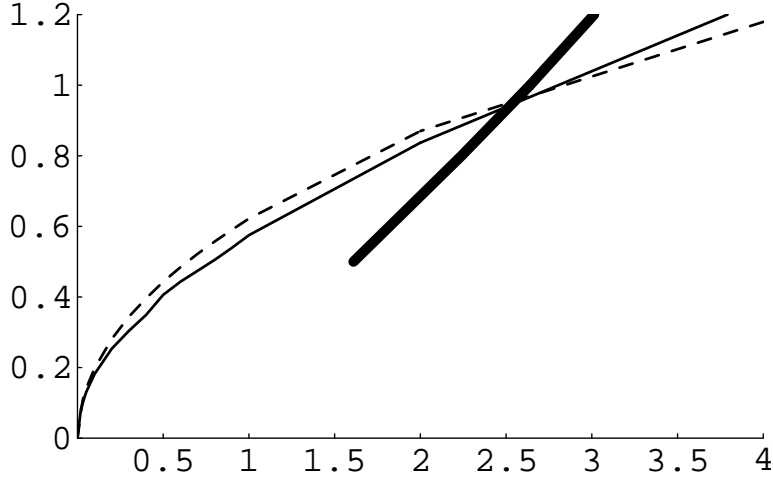}}
   \caption[FIG. \arabic{figure}.]{Resonance mass versus $\alpha$ for 
$p=3$ (solid line), $4$ (dotted line) and $5$ (bold line).}
\end{figure}
Figure 6 shows the resonance mass versus $\alpha$ for $p=3,4,5$. We classify 
them by the corresponding near-horizon brane configurations in ten or
eleven dimensions. The solid line corresponds to the D$3$-brane. The
dotted line corresponds to the D$4$ or M$5$-brane. The bold line 
corresponds to the D$5$-brane or an M$5$-brane intersection. It is
rather curious that the three lines appear to intersect at the same point,
at approximately $\alpha=2.688$ and $m/k=.976$.

\appendix
\section{$p=5$ as an exactly solvable model}

Consider the equations (\ref{Wave}) and (\ref{V}). Although this is not in
Schr\"{o}dinger form, for $p=5$ this equation is easily solvable for the
massive case. The solution for $\Theta$ is 
\be
\Theta=N(1+k|w|)^{1/2}\big[J_{-i\gamma}\Big(\frac{i\alpha}{1+k|w|}\Big) 
+A(\alpha)Y_{-i\gamma}\Big(\frac{i\alpha}{1+k|w|}\Big)],
\ee
where
\be
A(\alpha)=\frac{2 J_{-i\gamma}(i \alpha)+i\alpha[J_{-i\gamma+1}(i\alpha)-
J_{-i\gamma-1}(i\alpha)}{-2Y_{-i\gamma}(i\alpha)+i\alpha[Y_{-i\gamma-1}
(i\alpha)-Y_{-i\gamma+1}(i\alpha)]},\label{A5}
\ee
and $\gamma \equiv \sqrt{4(m/k)^2-1}$. Transforming back to the $z$
coordinate and $\psi$ wave function using (\ref{ztrans}) and
(\ref{wavetrans}) we obtain the wave function solution
\be
\psi(z)=N_0 \big[ J_{-i\gamma} (i\alpha\ e^{-k|z|/2}) +
A(\alpha) Y_{-i\gamma}(i \alpha\ e^{-k|z|/2}) \big].
\ee
For the massless graviton
\be
\psi(z)\sim e^{-k|z|/2}+A(\alpha) e^{k|z|/2},
\ee
and so the massless graviton wave function is not normalizable for nonzero
$B$ or $E$ fields.

As in the zero $B$-field case, there is a mass gap and so we
must have $m^2 \ge \frac{1}{4}k^2$ for the massive wave
functions. Expanding the Bessel functions for large $k|z|$ yields
\begin{eqnarray}
\psi(z) &=& N\frac{e^{\pi/2}}{\Gamma(1-i\gamma)} \Big(
[1+iA(\alpha)\rm{csch}(\pi\gamma)\cosh(\pi\gamma)]e^{iq|z|-i\gamma
\ln(\alpha/2)}-\\ \nonumber 
& & iA(\alpha)\rm{csch}(\pi\gamma)e^{-\pi}\frac{\Gamma(1-i\gamma)}
{\Gamma(1+i\gamma)}e^{-iq|z|+i\gamma \ln(\alpha/2)}\Big), 
\end{eqnarray}
where $q \equiv \sqrt{m^2-k^2/4}$. We numerically normalize the wave
function, and find the resonant mass by solving for the maximum of
$|\psi(z=0)|^2$ for a given $\alpha$. The resonant mass versus $\alpha$ is
plotted for $p=5$ as the bold line in Figure 6. With this exact solution
to the wave function, one can then find an analytic expression for how the
gravitational potential is modified by the presence of the constant,
background $B$ and $E$ fields.

Note that, for the case of zero $B$-field, the $p=5$ wave equation that
results from not dropping the "1" in the harmonic function $H$ is also
exactly solvable as a sum of Bessel functions \cite{poritz3}, using
the $w$ coordinate. This is of interest if one wishes to explore the
localization of gravity analytically in two-scalar domain-wall solutions
\cite{cvetic}.

\section{Acknowledgements}

I would like to thank M. Cveti\v{c}, G. Shiu, A. Karch, A. Naqvi and A. 
V\'{a}zquez-Poritz for useful discussions. I am especially grateful to
H. L\"{u} for extensive and very helpful discussions.

\end{document}